\newcommand{\beq}{\begin{equation}}
\newcommand{\eeq}{\end{equation}}
\newcommand{\beqa}{\begin{eqnarray}}
\newcommand{\eeqa}{\end{eqnarray}}

\documentstyle[aps,preprint,amsfonts,epsf]{revtex}
\begin{document}
\def\dfrac#1#2{{\displaystyle{#1\over#2}}}


\preprint{LA-UR-95-1638}

\title{Dynamical Properties of Quantum Hall Edge States}

\author { A.V.Balatsky$^{+}$*  and S.I.Matveenko *$^{+}$}
\address{$^{+}$ Theoretical Division, Los Alamos National Laboratory, Los
Alamos and,
 New Mexico 87545\\ $\ast$ Landau Institute for Theoretical Physics,
Moscow, Kosygina str., 2, Russia, 117940}

\date{April 5, 1995}

\maketitle

\begin{abstract}
We consider the dynamical properties of simple edge states in integer
($\nu = 1$) and fractional ($ \nu = 1/2m+1$) quantum Hall (QH)
liquids.  The influence of a time-dependent local perturbation on the
ground state is investigated. It is shown that the orthogonality
catastrophe occurs for the initial and final state overlap $|<i|f>|
\sim L^{-{1\over{2\nu}}({\delta\over{\pi}})^2}$ with the phase shift
$\delta$. The transition probability for the x-ray problem is also
found with the index, dependent on $\nu$. Optical experiments that
measure the x-ray response of the QH edge are discussed. We also
consider electrons tunneling from one dimensional Fermi liquid into a
QH fluid. For any filling fraction the tunneling
from a Fermi liquid to the QH edge is suppressed at low temperatures
and we find the nonlinear $I-V$ characteristics $I\sim V^{1/\nu}$.
\end{abstract}


\




The quantum Hall (QH) liquid is an incompressible state, where all bulk
excitations have finite energy gaps. As for any incompressible liquid,
droplet of water being the simplest example, the only low energy
excitations are surface modes. Upon quantizing these surface modes one
arrives at the picture of the simple propagating excitations on the
surface of the droplet. For the two dimensional QH liquid these
excitations represent the extended gapless edge excitations
\cite{hal,wen}. As was shown by Halperin, edge excitations in  the
integer QH (IQH) systems are described by a chiral 1D Fermi liquid
theory \cite{hal}. Similarly, Wen showed that  in the fractional QH
(FQH) state the edge excitations are described by a chiral Luttinger
liquid theory \cite{wen}.
 Within last few years  the transport in the chiral Luttinger liquid
was investigated in great detail, see for example \cite{wen,fisher}. We
argue that the  time dependent phenomena in this system deserve a
separate detailed investigation as well.

In this article few dynamical phenomena in the QH edge state will be
addressed.  The simplest question reflecting the nontrivial dynamics of
QH state is the orthogonality catastrophe \cite{Anderson}.  We
calculate the response of the QH edge states to local time-dependent
perturbations. And show that, for the case of sufficiently fast
switching on of a perturbation potential or slow switching off, the
orthogonality catastrophe takes place.  We find that the filling factor
$\nu$ enters into nontrivial exponents as well as in the phase shift
 $\delta = -\nu V_0/(2v)$,
where $V_0$ is the local potential, and $v$ is the velocity of the edge
mode (in 1D the density of states $N_0 = 1/v$). In the case of the
x-ray problem the transition rate is given by a power law of the
frequency near the threshold $W(\omega) \sim (\omega +E_0)^{\gamma}$
with power $\gamma = (1 - \delta /\pi)^{2}/\nu - 1$, where $E_0$ is the
deep core hole energy.

We also consider the tunneling via a point contact of the electron
from the metal into the QH edge. The strongly correlated nature of the
QH liquid with the composite boson order parameter, representing
binding of the flux with the fermions makes the excitations in the QH
states sufficiently different from the quasiparticles in a Fermi
liquid.  This results in the suppression of the tunneling conductance
at low temperature, bias voltage or driving frequency. We find $I \sim
V^{2/\nu +1}$ for the tunneling characteristics.

The orthogonality catastrophe and x-ray singularity were previously
considered for another 1-D system: The Luttinger liquid \cite{ll}.
There are qualitative differences between the Luttinger liquid and the
QH edge response to the x-ray transition. Firstly, the backscattering,
important for the x-ray response of a Luttinger liquid, is not present
in the QH system due to chiral nature of the state ( we do not
consider the role of disorder and the transition between opposite
edges).  Secondly, the presence of the fractionally charged
quasiparticles in the QH liquid leads to the ``fractionalization'' of
the phase shift with $\nu$ entering into all relevant exponents in the
problem. Although natural for the QH state, this nontrivial dependence
on the filling fraction (i.e.  fractional statistics and fractional
charge) can not be observed in the conventional Luttinger liquid.

We consider fractional states at odd integer $\nu^{-1}$. In this case
the edge state is known to be a single channel chiral Luttinger liquid
\cite{wen} and the approach, used in \cite{wen,fisher}, will be used.

{\em Local time-dependent perturbation}. The Hamiltonian of the edge
waves is given by \cite{wen}
\beq
H = \int dx \frac{\pi}{\nu} \rho^2 v = \frac{2\pi }{\nu} v \sum_{k>0}\rho_{k}
\rho_{-k}
\label{e1}
\eeq
where $\rho$ is the one dimensional density, $v = cE/B$ is the
velocity of electron drift near the edge, $E$ is the electric field of
the potential well, $B$ is the magnetic field. The Hamiltonian
Eq.(\ref{e1}) is quantized in terms of the $U(1)$ Kac-Moody algebra
\cite{wen} and we have
\beq
[\rho_k, \rho_{k\prime} ] = \frac{\nu}{2\pi} k \delta_{k + k\prime, 0} \ \
[H, \rho_{k}] = v k \rho_{k}
\label{e2}
\eeq

A similar algebra (with change $\nu \rightarrow L$) is known in the
Tomonaga model\cite{Tomonaga}. The electron operator is written in
terms of the field $\varphi$ as $\Psi \propto \exp{i\varphi /\nu}$ with the
commutator
\beq
[\varphi (x), \varphi(y)] = i\pi \nu sign (y-x)
\label{e4}
\eeq
The density operator $\rho (x)$ is equal to $\rho (x) = \partial_x \varphi
/(2\pi)$. The action in terms of the chiral boson field $\varphi$ is
\beq
S = \frac{1}{4\pi \nu} \partial_x \varphi (\partial_{t} \varphi +
v\partial_x \varphi )
\label{e5}
\eeq
The interaction of the system with an external time-dependent
perturbation may be written in the form
\beq
H_{int} = \int dx V(x,t) \rho (x,t)
\label{e6}
\eeq
Consider the following expressions for $V(x,t)$, for different time
dependence of the potential (spatial from of $V(x)$ remains the same
for all cases):
\beq
V(x,t) = V_0 \theta (t) (1 - \exp({-t/\tau})) \delta (x)
\label{e7}
\eeq
\beq
V(x,t) = V_0 \theta (t)  \exp({-t/\tau}) \delta (x)
\label{ee7}
\eeq
\beq
V(x,t) = V_0 \theta (t) \exp({i\Omega t - \epsilon t}) (1 -
\exp({-t/\tau})) \delta (x),\; \epsilon \rightarrow +0
\label{e8}
\eeq
In all cases the potential is nonzero only for positive times due to step
function in Eqs.(6-8)  and there is no violation of causality.

The matrix element between initial $\vert i >$ and final
$\vert f > = T_t \exp[i\int_0^t H_{int}dt] \vert i >$  states in the limit
$t \rightarrow \infty$ is given by the following:
\beq
<f\vert i>={\int D\varphi \exp[i\int dxdt(S + V(x,t) \rho (x,t))]\over{\int
D\varphi
\exp[i\int dxdtS]}} .
\label{e9}
\eeq

The integration in Eq.(\ref{e9}) is trivial and we have as a result
\beq
|<f \vert i >| = \left[ \frac{1 + (L/2\pi \tau v)^2}{1  +(1/p_{0}\tau v)^{2}}
\right]^{-{1\over{4\nu}} {\delta^2\over{\pi^2}}},
\label{e10}
\eeq
for the switching on  the  potential, Eq. (\ref{e7}), and
\beq
|<f \vert i >| = \left[\frac{1+(p_{0}\tau v)^{2}}
{1 + (2\pi \tau v/L)^{2}} \right]^{-{1\over{4\nu}} {\delta^2\over{\pi^2}}}, \
\delta = -{\nu V_0\over{2v}}
\label{ee11}
\eeq
for  switching off  the potential, Eq.(\ref{ee7}), where $L$ is
the system length and  $p_0$ is the ultraviolet momentum cut off (of the
order of the inverse magnetic length).

In the case of slow switching off ($\tau v \gg L$) the potential or
fast switching on ($\tau v \ll L$) we have,
\beq
|<f \vert i >|  \propto L^
{-\frac{1}{2 \nu}\frac{\delta^2}{\pi^2}} .
\label{e12}
\eeq
The time dependence of the matrix element   at  time
$t \gg \tau$ for the potential (\ref{e7}) is easily found:
\beq
|<f (t) \vert i (0) >|  \propto t^
{-\frac{1}{ \nu}\frac{\delta^2}{\pi^2}} ,
\label{ee13}
\eeq
These results are exact in the infrared limit of the QHE
for which the effective Hamiltonian Eq.(\ref{e5}) was derived.

In the case of the periodic  potential Eq.(\ref{e8})  $|<f\vert i>|\propto
\exp{-\Omega/\epsilon} \rightarrow 0$, for $\Omega / v < p_{0}$ and
\beq
|<f \vert i >|  \propto e^
{-\frac{1}{2 \nu}\frac{\delta^2}{\pi^2}X}
\label{ee14}
\eeq
for $\Omega > p_{0} v$, where
$ X = 2 \log c + 2(1-c)/c$ in the limit $\Omega \tau \ll 1$,
$ X = \frac{(c-1)^{2} (c+2)}{3c^{2} (\Omega \tau)^{2}}$ if
$\Omega \tau \gg 1$, and  $ c = 1 - p_{0} v /\Omega$.

The edge states support the nondecaying ``particle-hole" pair - the
edge wave. This propagating mode with $\chi"(q, \omega)\propto
\delta(\omega - vk)$ ($\chi(q, \omega)$ being the density-density correlator)
is sufficient to generate the orthogonality in
1D.  This would not be the case in higher dimensions, where the
existence of the particle-hole continuum, e.g. $\chi"(q,
\omega)\propto (\omega/vq) \Theta(\omega - vq)$, with large phase space
is required to get orthogonality catastrophe.

Note also that our result Eq.(\ref{e12}) {\em does not} reproduce the
Fermi liquid result at $\nu = 1$. This is the consequence of the
chiral nature of the excitations even at $\nu = 1$. To recover the
Fermi liquid result, consider the edge of the Fermi liquid state (if
one with only edge gapless excitations would exist) as the product of
two chiral Slater determinants, each one corresponding to the right
and left moving particles. Then Eq.(\ref{e10}) at $\nu =1$ represents
the response of the chiral branch of the Fermi liquid. To get the full
answer for the total overlap one has to multiply $<f|i>_L<f|i>_R =
<f|i>^2$ and one recovers the standard Fermi liquid result by taking
square of the Eq(\ref{e12}) for $\nu = 1$.

Next, we calculate the transition probability for the x-ray problem.
Because of the gapped spectrum in the bulk of QH fluid the only
gapless excitations to contribute to the x-ray response near frequency
threshold are edge excitations.  Imagine the x-ray process, creating
the deep hole near the edge in a QH sample. Creation of the hole at
energy $E_0$ will lead to the Coulomb potential of the hole and to
 injection of  an electron into the edge of the sample. As
is common for the x-ray problem, the response of the edge states will
contain two contributions: the orthogonality of the initial and final
state Eq(\ref{e10}) and the distortion of the ground state due to the
extra electron. Following the classical paper of Schotte and Schotte
\cite{SS}, we first calculate the propagator:
\beqa
F(t) = <i|e^{i \int^t H_i dt} \Psi e^{-i\int^t H_f dt} \Psi^{\dag}|i>
\eeqa
with $H_i$ being the unperturbed Hamiltonian Eq(\ref{e1}), $H_f = H_i
+ H_{int} $ is the final state Hamiltonian, and $\Psi$ is the QH
liquid electron operator. We assume here that the deep core hole was
created in the absorption and is localized. For the case of scatterer
in the valence band recoil effects will change our results. Recoil
also can be addressed within this bosonization scheme.  Using the bosonic
algebra we find
\beqa
F(t) \sim t^{- {1\over{\nu}} (1-{\delta\over{\pi}})^2} .
\eeqa
{}From this result the transition probability near the frequency edge
$W(\omega) \sim Im \ i/\pi \int_0^{\infty} F(t) e^{i(\omega +E_0)} dt$
becomes:
\beqa
W(\omega) \sim (\omega + E_0)^{{1\over{\nu}}(1-{\delta\over{\pi}})^2 - 1} .
\label{e21}
\eeqa
In the absence of perturbation  $V$ a similar formula was
obtained for the electron density of states \cite{wen2} $N(\omega )
\propto \int dk \sum_{n} \vert <n\vert \Psi_{k} \vert 0>\vert^{2}
\delta (\omega - \omega_{n}) \propto \omega^{-1 + 1/\nu}$.  The result
depends on the potential whose phase shift $\delta$ can be either
positive or negative. One interesting outcome of this result is that
even for fractional QH state the transition probability near the threshold
will behave differently for attractive and repulsive interactions. Take
for example the unitary phase shift $\delta = \pm \pi/2$ for attractive
and repulsive interaction respectively.  Then $W(\omega) \sim
\omega^{1/(4\nu) - 1} \ \ (attraction); \ \ W(\omega) \sim
\omega^{9/(4\nu)-1} \ \ (repulsion)$. And we find even for the $\nu = 1/3$
state that an attractive potential leads to an  enhanced transition rate
near the threshold: $W(\omega) \sim \omega^{-1/4}$.

To be able to measure the x-ray response of the QH edge, one has to
excite a deep hole in the valence-conduction band transition in a
semiconductor (with typical bandwidth of the order of 1 eV) and be
able to tune the laser to within cyclotron frequency (1-5
meV) to excite only the lowest Landau level and avoid any interlevel
excitations. Although the gapped excitations might be excited in this
process, near the threshold frequency, only the soft gapless edge
excitations  provide the dominant contribution. In the most
interesting case of the  FQH edge the gaps will be given by the
many-body  gaps and are somewhat smaller, although of the same
magnitude.

There is one interesting aspect of the orthogonality catastrophe in the
fractional QH state which makes this problem qualitatively different
from the similar problem in a Fermi liquid. In principle on can imagine
the {\em ``decay''} of the electron, injected into the edge, on to
fractionally charged quasiparticles, which will provide an extra source
of the orthogonalization. However, the simple edge model, used here,
does not allow us to address this question. Indeed, in the simple edge
model the velocity of the charge on the edge is independent of the
charge and therefore the  electron will propagate along the edge as a wave
packet, which can be considered as made from the quasiparticles. It is
only the inclusion of the nonlinear terms and interaction between
quasiparticles that  might induce the decay of the wave packet.

We can consider the motion of the extra electron taking into account
the perturbation potential $H_{int}$. The  matrix element of decay of the
electron into $1/\nu$ quasiparticles at points $x_{i}$ at time $t$ has
the form,
\beqa
\tilde{F}(t) = <i|\prod_{j}^{1/\nu} e^{i\varphi (x_{j},t)}  e^{i \int^t H_{int}
dt} e^{-i\varphi (0,0)}|i> \propto \nonumber \\
(vt)^{\frac{1}{\nu}\frac{\delta}{\pi} (1 - \frac{\delta}{\pi})}
\frac{1}{\prod_{j}^{1/\nu} (x_{j} - vt)^{1 - \delta /\pi} x^{\delta
/\pi}_{j}} .
\label{ee1}
\eeqa
We see that the function $\tilde{F}(t)$ has two maxima : at point
$x_{j}=0$ and $x_{j} = vt$. These two peaks have a simple physical
origin, the first one near $x=0$ comes from the screening electron
density, induced by the attractive hole potential, while the second
one is coming form the propagating mode which  moves the electron
density along the edge with velocity $v$.  For the case of a strong
potential ($ \delta = \pi /2$) the probabilities to find the
quasiparticles at these points are equal to $(vtp_{0})^{- 1/(2\nu)}$.
Therefore, we assume that the average values of the electric charge at
$x=0$ and at $x = vt$ are the same and equal to $e/2$.  Due to the
orthogonality catastrophe the matrix element $\tilde{F}(t) \rightarrow
0$ and this effect may be observable only for $vt < p_{0}^{-1}$.

{\em Tunneling from metal into QH edge}. Next we consider a tunnel
junction or point contact between a 1D Fermi liquid and a QH system.
Experimentally this situation can be realized by attaching the contact
to the edge of the Hall bar.

 The Euclidean action of the QH states in terms of a chiral boson field
$\varphi$ follows from Eq. (\ref{e5}) with the substitution $t \rightarrow
i\tau$.
The noninteracting Fermi liquid can be considered as a superposition of
two chiral Fermi liquids (which are described by the action
Eq.(\ref{e5}) at $\nu = 1$) with opposite Fermi velocities. The
corresponding actions differ by the sign of the $i\partial_{\tau} \varphi$
term:
\beq
S_F = S_{\nu = 1} (\Psi_L ) + S_{\nu = 1} (\Psi_R ) .
\label{e14}
\eeq
The tunneling Hamiltonian has the form
\beq
H_{TUN} =  (t_1 \Psi_R \Psi^{\dag}  + t_2 \Psi_L \Psi^{\dag} )\delta (x) + h.c.
\label{e15}
\eeq
Where $\Psi_{L,R}$ are the Fermi liquid left(right) electron
operators, $\Psi$ is the electron operator in QH liquid. We do not
take into account the spin of the electrons, as we consider only states with
spins
parallel to the magnetic field $B$. The opposite spin
component can tunnel in the presence of the spin scattering in the
contact. Inclusion of this term will simply modify $H_{TUN}$.

In terms of corresponding boson fields $\varphi_{L}(x,\tau)$,
$\varphi_{R}(x,\tau)$, $\varphi (x,\tau)$
  ($\Psi_{R,L} \propto \exp{\pm i \varphi_{R,L}}$) we have for the tunneling
Hamiltonian
\beqa
H_{TUN}=t_{1}\cos(\varphi_{R}(0,\tau)-\varphi(0,\tau)/\nu)+ \nonumber\\
t_{2}\cos(\varphi_{L}(0,\tau)+\varphi(0,\tau)/\nu) .
\label{e16}
\eeqa
Since the perturbation terms act at the one point $x=0$ it is useful
to perform a partial trace in the partition function and integrate in
$\varphi ,\varphi_{R}, \varphi_{L}$ for all $x\neq 0$. If we define $\varphi
(0,
\tau) = \varphi (\tau ), \varphi_{R,L} (0, \tau) = \varphi_{R,L} (\tau )$ then
the actions $S_{FQH}, S_{R,L} $ are minimized when
\beqa
\varphi (x, \omega_{n}) = &\varphi (\omega_{n}), sign(\omega_n x) \leq
0;\nonumber \\
                    = &\varphi (\omega_{n}) \exp{(-\omega_{n}x)},
sign(\omega_n x)>0
\label{e17}
\eeqa
and similar expressions for $\varphi_{R}, \varphi_{L}$. The resulting action is
\beqa
S_{eff} = \frac{1}{4\pi}[{1\over{\nu}} \sum_{n}\vert \omega_{n}\vert \vert
\varphi (\omega_{n})\vert^{2} + \nonumber\\
 \sum_{n}\vert \omega_{n}\vert \vert \varphi_{R} (\omega_{n})\vert^{2} +
\sum_{n}\vert \omega_{n}\vert \vert \varphi_{L} (\omega_{n})\vert^{2}]
\label{e18}
\eeqa

In the lowest order perturbation expansion we obtain the
renormalization group (RG) equations for $t_1 , t_2 $ upon rescaling $\tau
\rightarrow l \tau$
\beq
{\partial t_{1,2}\over{\partial \ln l}} = t_{1,2}/2 ( 1 - \nu^{-1}) .
\label{e20}
\eeq
Both $t_{1}$ and $t_{2}$ terms flow to zero, indicating that at low
energies the junction is insulating. Note that we consider only simple
edge states with $\nu \leq 1$. This result can be interpreted as the
reflection of the zero overlap between the ground state quasiparticles
in the QH liquid with the Fermi liquid quasiparticles, excluding $\nu = 1$
case, where tunneling is not suppressed.

If we cut off the RG flow at some infrared energy scale $\Lambda$
(corresponding to temperature  $T$,  voltage $V$ or frequency ), we obtain
\beq
t_{1,2} \propto \Lambda^{1/2(\nu^{-1} - 1)}
\label{e22}
\eeq
For $\nu\leq 1$ the conductance of the contact is defined by the largest
of $t_{1,2}$ for small $\Lambda$. In this case the conductance of the
junction is:
\beq
G \sim
t_{i}^{2} \sim V^{\nu^{-1} - 1}
\label{e23}
\eeq
Therefore we find the power law $I-V$ characteristics, $ I \sim
V^{\nu^{-1}}$. The nonlinearity in the $I-V$ characteristic is
quite strong and, for example, for $\nu = 1/3$ we find $I\sim V^3$,
which might be detectable experimentally.


{\em In conclusion}, we investigated the influence of time-dependent
local perturbations on the ground state of the FQHE state and have
calculated the matrix element between the initial and the final ($t
\rightarrow \infty$) states. In the limit of sufficiently slow
switching off the potential or fast switching on the orthogonality catastrophe
occurs. The x-ray problem on the edge of the FQHE state, where the
existence of the gapless excitations changes the response near
singularity, is solved. The transition probability for the absorption
is $W(\omega) \sim
\omega^{{1\over{\nu}}(1-{\delta\over{\pi}})^2 - 1}$. We find that in
these problems the phase shift is ``fractionalized" $\delta = -{\nu
V_0\over{2 v}}$ with filling fraction $\nu$ entering into all relevant
exponents. We find that for the $\nu = 1/3$ fractional QH state the
transition probability is enhanced(suppressed) near the threshold for
attractive(repulsive) unitary potential, whereas for any lower filling
it will  always be  suppressed. This is not the case for the lower
filling fractions. For the weak potential with $\delta \rightarrow 0$,
on the other hand, the transition rate is suppressed for any filling
and for both signs of interaction. This situation is {\em
qualitatively} different from the case of Fermi liquid (Eq.(17) with
$\nu = 1$), where even in the case of weak potential the sign of the
phase shift matters and the transition rate is suppressed(enhanced) for
repulsion(attraction). A possible experiment, which  probe the x-ray
response of the edge might be based on the valence-conduction band
transition due to semiconducting laser light absorption, creating a deep hole
and electron near the edge. The absorption coefficient will be proportional to
the probability $W(\omega)$ Eq. (\ref{e21}). To maximize the absorption at the
edges the sample can be covered with only the edges  being lightened with
laser. If the
frequency of the transition is tuned so that gapped bulk excitations
are not excited, the dominant response will come from the excited edge
states.

We considered the tunnel junction between FQHE system and 1D Fermi
liquid as well.  We find that for any filling $\nu \leq 1$ the
tunneling junction is insulating in the limit $T=0$ and the nonlinear
conductance of the junctions leads to the $I\sim V^{1/\nu}$
tunneling characteristics. We interpret the insulating behavior of the
junction as the result of the orthogonality of the electron and the
quasiparticles in the QH liquid, except the IQHE case.

Investigation of the dynamical response of the QH states
might be an interesting experimental field. We propose
to measure the nonlinear tunneling conductance from the QH edge into the
metal. If detectable at small bias at low temperature, the $I-V$
characteristics of such a junction will be strongly nonlinear ($I \sim
V^3$ for $\nu = 1/3$). The nontrivial dynamics of the edge states can
be probed in the experiments with the x-ray transitions in the region
near the edge. The x-ray transition will be suppressed or enhanced at
the threshold, as described above.

We are grateful to K. Bedell for comments and careful reading of the
manuscript.  A.B.  would like to thank Prof. H. Jiang for discussion on
the possible x-ray transitions experiments in QH liquid. S.M. wishes to
thank T-11 and Strongly Correlated Electron Theory program at CMS for
hospitality. A.B. thanks M.P.A. Fisher for discussion and for pointing
out the mistake in the first version of this paper.
 This work has been supported in part by the US Department of Energy.

\end{document}